\documentclass{ws-p9-75x6-50}

\begin{document}

\title{Minisuperspace Model for Revised 
Canonical Quantum Gravity}
\author{Giovanni Montani }

\address{ICRA, Dipartimento di Fisica (G9), Universit\'a
di Roma ``La Sapienza", P.le Aldo Moro 5, Roma 00185, Italy\\
E-mail: montani@icra.it}

\maketitle

\abstracts{
We present a reformulation of the canonical
quantization of gravity, as referred to the minisuperspace;
the new approach is based on fixing a Gaussian 
(or synchronous) reference frame 
and then quantizing the system via the reconstruction
of a suitable constraint; then the quantum dynamics
is re-stated in a generic coordinates system
and it becomes dependent on the lapse function.\\
The analysis follows a parallelism
with the case of the non-relativistic particle and leads to
the minisuperspace implementation of the so-called
{\em kinematical action} as proposed in \cite{M02}
(here almost coinciding also with the approach
presented in \cite{KT91}).\\ 
The new constraint leads to a schr\"odinger equation
for the system. i.e. to non-vanishing eigenvalues for the
super-Hamiltonian operator; the physical
interpretation of this feature relies on the appearance
of a ``dust fluid'' (non-positive definite) energy density,
i.e. a kind of ``materialization'' of the reference frame.\\ 
As an example of minisuperspace model, we consider
a Bianchi type IX Universe, for which some
dynamical implications of the revised canonical quantum
gravity are discussed.
We also show how, on the
classical limit, the presence of the dust fluid
can have relevant cosmological issues.\\
Finally we upgrade our analysis by its
extension to the generic cosmological solution, which
is performed in the so-called long-wavelength
approximation. In fact, near the Big-Bang,
we can neglect the spatial
gradients of the dynamical variables and 
arrive to implement, in each space point, the same
minisuperspace paradigm valid for the Bianchi IX model.}

\section{Introduction}

One of the most peculiar and puzzling features that
the canonical quantum method has to face
when it is applied to the 
gravitational field consists of a vanishing
Hamiltonian function. Such a dynamical constraint
reflects the invariance of the theory under
infinitesimal time displacements and results into
the non evolutive character singled
out by the Wheeler-DeWitt equation (WDE) \cite{D67} 
(see also \cite{K81}). Indeed, via the
Arnowitt-Deser-Misner (ADM) (3+1)-splitting
of the space-time \cite{ADM62,MTW73},
the quantum information about the gravitational field is
provided by a wave functional taken on a whole class of
3-geometries; in fact, the theory should be invariant 
under spatial coordinates re-parameterizations
which are equivalent to the gauge  symmetry observed in
non-Abelian theories and is ensured 
requiring that the wave functional be annihilated
by the super-momentum operator, i.e. $\hat{H}_i\Psi =0$).
Such equations restricts to a class of 3-geometries 
the dynamical variable on which the wave function is taken
and then the canonical quantum dynamics is provided by
requiring that also the super-Hamiltonian operator annihilates
the system states,
i.e. $\hat{H}\Psi =0$;
with respect to such a WDE equation
we stress the following shortcomings features \cite{D97}:\\
i) The WDE is a covariant quantum theory, 
but $\Psi$ does not depend on the lapse function $N$ and
the shift vector $N^i$
(because of the two constraints associated with the
vanishing of their conjugate momenta) and 
we have information only on the 3-geometries.\\
ii) The wave functional takes the same value on each
spatial hypersurface of the slicing and, therefore,
no real evolution takes place
(for a review about the problem of ``time''
in canonical quantum gravity see \cite{I92,R91}).\\
iii) The quantum dynamical equation,
in view of its hyperbolic structure.
prevents a general prescription to arrange the
space of the solutions into an Hilbert space
and consequently no
{\em probability notion} is naturally defined.

Over the years, many approaches have been presented
in order to construct an internal physical clock for
the quantum system and then to achieve
the Hilbert space structure. In a recent
work \cite{M02}, it has been proposed a correlation
between the ``frozen formalism''
of the WDE and the ambiguity in
developing a (3+1)-slicing of a quantum
space-time; in fact, it makes no precise sense,  
in a quantum picture, to 
speak of spatial hypersurfaces
(or of a time-like normal)
because such notions can be recognized,
at most in terms of expectations values.
Thus we infer that 
to implement a straightforward canonical quantization
of the slicing can be responsible for a dynamics
without evolution.

As a solution to the above ambiguity,
in \cite{M02}, was proposed to include the so-called
{\em kinematical action} in treating the gravitational
problem, as done for a quantum field on a fixed
metric background.\\ 
Such a procedure is essentially equivalent
to fix the reference frame before quantizing the
gravitational field; 
This new approach finds its
physical interpretation in the appearance of a real
clock, consisting of a dust fluid; indeed the 
equations of motion associated to the 
kinematical action can be rewritten,
under suitable hypotheses, 
as a dust fluid dynamics \cite{M02}.\\
The aim of \cite{M02} is to argue
that quantizing via a
canonical method, the 3-geometries, it requires
the existence of a ``clock fluid'' that makes physical
the slicing.\\

Here we consider a minisuperspace cosmological model in a
Gaussian  (or synchronous) reference and quantize it in close
analogy with the non-relativistic particle \cite{K81}; 
as a result, we get the same quantum dynamics provided
for this case by the kinematical term
(as well as, by the, here almost
overlapping, approach presented in \cite{KT91}).\\
The quantization is then referred to a generic
reference frame and we achieve
a dynamical picture at all
equivalent to that one proposed in \cite{M02};
such a coincidence of two different approaches confirms the
necessity of fixing the reference frame before
the quantization of a ($3+1$)-slicing. 

Then we show how it is possible for our system,
to construct a natural Hilbert space and how,
the evolution of the wave function
becomes interpretable 
via a ``dust fluid'' of reference;
finally we outline
some relevant features concerning the semiclassical limit
of the model.\\

As minisuperspace model we consider the Bianchi type IX
cosmology and then we extend the analysis
to a generic inhomogeneous
cosmological solution; in fact we show how the generic case,
close enough to the cosmological singularity,
can is reduced to a point-like minisuperspace dynamics. 
This feature is due to the dynamical decoupling of the space
points which takes place when the Universe volume approaches
zero near the Big-Bang.
From a quantum dynamical point of view, we will deal with
the generic cosmological solution in the limit of the so-called
long-wavelength approximation; such an inhomogeneous extension
provides with degree of generality  
to all the results derived for the Bianchi IX model. 

In section $2$ is discussed the quantization of the
non-relativistic parameterized particle, regarded as
the prototype for building up in section $3$ an
appropriate analysis of the Bianchi IX quantum
dynamics, as viewed in a Gaussian frame. In section
$4$ is developed the classical limit of the revised
minisuperspace quantization and some relevant cosmological
implications of the outcoming picture are presented.\\

Section $5$ is devoted to upgrading our previous
analysis, by showing how, under well-grounded assumptions,
it can be extended to the generic cosmological solution;
by other words, we outline that, in each space point of
a generic inhomogeneous Universe, takes place, independently,
the same minisuperspace picture characterizing a
Bianchi IX cosmology. In section $6$ concluding remarks
follows.

\section{Parameterized particle}

We start by reviewing the case of the one-dimensional
non-relativistic (parametrized) particle, whose
action reads

\begin{equation}
S=\int \{ p\dot{q} - h(p, q)\} dt \, ,
\, ,
\label{a}
\end{equation}

where $t$ denotes the Newton time and $h$ the
Hamiltonian function. In order to quantize this
system, we parameterize the Newton time as $t=t(\tau )$,
which leads to have

\begin{equation}
S=\int \{ p\frac{dq}{d\tau } - h(p, q)\frac{dt}{d\tau }\} d\tau
\, .
\label{b}
\end{equation}

Now we set $p_0\equiv -h$ and add this relation to the
new action by a Lagrangian multiplier $\lambda $, i.e.

\begin{equation}
S=\int \{ p\frac{dq}{d\tau } + p_0\frac{dt}{d\tau } -
\bar{h}(p, q, p_0, \lambda )\} d\tau
\, \quad \bar{h}\equiv \lambda (h + p_0)
\, .
\label{c}
\end{equation}

By varying this action with respect to $p$ and $q$,
we get the Hamilton equations

\begin{equation}
\frac{dq}{d\tau } = \lambda
\frac{\partial h}{\partial p}
\, \quad 
\frac{dp}{d\tau } = -\lambda \frac{\partial h}{\partial q},
\, , 
\label{he}
\end{equation}

while the variations of $p_0$ and $t$ yield

\begin{equation}
\frac{dt}{d\tau } = \lambda 
\, \quad 
\frac{dp_0}{d\tau } = 0 
\, . 
\label{he1}
\end{equation}

All together, these equations describe the same Newton
dynamics with the energy as constant of the motion.
But now, by varying $\lambda$, we get the (desired)
constraint $h + p_0 = 0$, which, in terms of the
operators $\hat{p}_0 = -i\hbar \partial _t$
and $\hat{h}$,
provides the Schr\"odinger equation

\begin{equation}
i\hbar \partial _t \psi = \hat{h}\psi 
\label{sch}
\end{equation}

for the system
state function $\psi (t, q)$.
Finally we remark that, when retaining
the relation
$dt/d\tau = \lambda$, we are able to write the
wave equation in the parametric time as

\begin{equation}
i\hbar \partial _{\tau } \psi (\tau , q) = \lambda (\tau ) 
\hat{h}\psi (\tau , q) 
\, ; 
\label{d}
\end{equation}

where the function
$\lambda (\tau )$ is to be specified for
completing the dynamical scheme.

\section{Quantization of the Bianchi IX model}

Now we implement this same method of quantization
in the minisuperspace associated with an
homogeneous cosmological model of the type IX.\\
The Bianchi IX model is the most general one
(together with type VIII) 
allowed by the homogeneity constraint
and is described via a line element of the form
\cite{MTW73,LL75}

\begin{equation}
ds^2 = -N(t)^2dt^2 + \frac{R(t)^2}{6\pi }\left(
e^{2(\beta (t)}\right) _{ij} \sigma ^i(x^l)\sigma ^j(x^l)
\, \quad i,j,l=1,2,3 
\, , 
\label{e}
\end{equation}

where we take the diagonal form

\begin{equation}
\left( e^{2\beta (t)}\right) _{ij} =
diag. \left\{ e^{2(\beta _+ + \sqrt{3}\beta _-)}\; , \;
e^{2(\beta _+ - \sqrt{3}\beta _-)}\; , \; e^{-2\beta _+}\right\}
\,  
\label{f}
\end{equation}

and the 1-forms $\sigma ^i(x^l)$,
to which is associated the
Lie algebra of the isometries, read explicitly as 

\begin{equation}
\sigma ^1 = \cos \chi d\theta + \sin \chi \sin \theta d\varphi
\; , \;
\sigma ^2 = \sin \chi d\theta - \cos \chi \sin \theta d\varphi
\; , \;
\sigma ^3 = d\chi + \cos \theta d\varphi
\, ,
\label{g}
\end{equation}

with
$0\le \varphi < 2\pi$, $0\le \theta < \pi$ and $0\le \chi < 4\pi$.\\
By adopting these Misner-like variables \cite{M69},
we separate the isotropic (volume) expansion
of the Universe,
represented by the function $R(t)$, 
from its anisotropies, which are described
through the degrees
of freedom $\beta _+(t)$ and $\beta _-(t)$. 

The very early Universe evolution was characterized
by a thermal bath containing all the fundamental particles
and since most of the species were
described by an ultrarelativistic
equation of state, then we include into the problem
a phenomenological energy density
$\rho _{ur} = \mu ^2/R^4\; , \; \mu = const.$.
Furthermore, the idea that the Universe underwent an
inflationary scenario, leads us
to involve in the dynamics
a real self-interacting scalar field $\phi$; the 
associated ``finite temperature
potential'' $V_{\cal T}(\phi )$
(here ${\cal T}$ denotes the Universe temperature)
can be taken in
the Coleman-Weinberg form \cite{CW73} 

\begin{equation}
V_{\cal T}(\phi ) = \frac{B\sigma ^4}{2h^3c^3} +
B\frac{\phi ^4}{hc}\left[ \ln
\left( \frac{l_{Pl}\phi ^2}{\sigma ^2}\right) - \frac{1}{2}\right] + 
\frac{1}{2}{m_T}^2\phi ^2 \, \quad
m_T = \sqrt{\lambda T^2 - m^2} \, \quad
(m, \lambda ) = const. \, ;
\, , 
\label{CW}
\end{equation}

here $B\sim \mathcal{O}(10^{-3})$ and
$\sigma \sim \mathcal{O}(10^{14})GeV$ denotes the
scale of the transition phase.\\ 
In what follows, we regard
the Universe temperature as a
function of the variable $R$, i.e.
${\cal T}(R) = {\cal T}^*/R \; , \; {\cal T}^* = const.$.

The evolution of the cosmological model so obtained,
is summarized, in a Gaussian  reference
($N=c\; , \;  t\rightarrow T$), by the action

\begin{equation}
S_{IX}  = \int \left\{
p_R\frac{dR}{dT} + 
p_+\frac{d\beta _+}{dT} + 
p_-\frac{d\beta _-}{dT} +
p_{\phi }\frac{d\phi }{dT} -
{\cal H}(R, \beta _{\pm }, \phi , p_R, p_{\pm }, p_{\phi })
\right\} dT
\, , 
\label{h}
\end{equation}

where all the $p$`s denote the conjugate momenta to the
respective dynamical variables, and the
Hamiltonian function takes the form

\begin{equation} 
{\cal H} =
\sqrt{\frac{3\pi }{2}}\left\{ \frac{l_{Pl}^2}{2\hbar } \left[
-\frac{p_{R}^2}{R} +
\frac{1}{R^3}\left( p_+^2 + p_-^2 \right) \right] + 
\frac{3c^2}{8\pi } p_{\phi }^2 +
U(R,\; \phi ) \right\}
\, ,
\label{i}
\end{equation}

with the potential term 

\begin{equation} 
U(R,\; \phi ) \equiv 
\frac{\mu ^2}{R} + \frac{\hbar c}{l_{Pl}^2}R\left(
V(\beta _{\pm }) - 1\right) + \frac{32}{18}R^3
V_{\cal T}(\phi )
\, . 
\label{l}
\end{equation}

The characterization of this dynamical system is completed
by specifying the form of $V(\beta _{\pm })$
as \cite{M69,MTW73}:

\begin{equation} 
V(\beta _{\pm }) = \frac{1}{3}e^{-8\beta _+} -
\frac{4}{3}e^{-2\beta _+} \cosh (2\sqrt{3}\beta _-) + 1 +
\frac{2}{3}e^{4\beta _+} \left( \cosh (4\sqrt{3}\beta _-) - 1
\right) \, \quad V(0,\; 0) = 0 
\, .
\label{m}
\end{equation}

Now we quantize this system
in close analogy with the case of the
non-relativistic particle,
by re-parameterizing the Gaussian
time as follows $T = T(t)$; then, by the position
${\cal H} = -p_T$ (added to the action via a Lagrangian
multiplier $\Lambda (t)$), we get 

\begin{equation}
S_{IX}  = \int \left\{
p_R\frac{dR}{dt} + 
p_+\frac{d\beta _+}{dt} + 
p_-\frac{d\beta _-}{dt} +
p_{\phi }\frac{d\phi }{dt} + p_T\frac{dT}{dt} - 
\Lambda (t)(p_T + {\cal H})
\right\} dt
\, . 
\label{n}
\end{equation}

We stress how the variation of this
action with respect to
$T$ yields the key relation  $dT/dt = \Lambda (t)$ and,
by comparing it with (\ref{e}), it comes out that
$\Lambda (t) = \pm N(t)$; hence,
the choice of the positive root
for $\Lambda$, allows to rewrite
(\ref{n}) in the same form
as we would have obtained applying,
to the present case,
the general method discussed in \cite{M02}
(which is based on the use of the kinematical action), i.e.

\begin{equation}
S_{IX}  = \int \left\{
p_R\frac{dR}{dt} + 
p_+\frac{d\beta _+}{dt} + 
p_-\frac{d\beta _-}{dt} +
p_{\phi }\frac{d\phi }{dt} + p_T\frac{dT}{dt} - 
N(t)(p_T + {\cal H})
\right\} dt
\, . 
\label{o}
\end{equation}

In both the cases (\ref{n}) and (\ref{o}), as soon as we
implement the Hamiltonian constraint on a quantum level,
we are lead to a Schr\"odinger equation of the form
(the canonical quantization is based
on replacing the momentum
$p_*$, conjugate to the generic variable
$x^*$, by the operator
$\hat{p}_*= -i\hbar \partial _{x^*}$
and taking the Hamiltonian operator constraint
on the state function $\psi $)

\begin{equation} 
i\hbar \partial _T \psi = 
\hat{{\cal H}} \psi =
\sqrt{\frac{3\pi }{2}}\left\{ \frac{l_{Pl}^2\hbar }{2} \left[
\partial _R\left( \frac{1}{R} \right) \partial _R - 
\frac{1}{R^3}\left( \partial _+^2 + \partial _-^2 \right)
\right] - 
\frac{3c^2}{8\pi } \partial _{\phi }^2 +
U(R,\; \phi ) \right\} \psi 
\, , 
\label{p}
\end{equation}

with $\psi = \psi (T, \; R, \; \beta _{\pm }, \; \phi )$.

It is worth stressing that, in analogy to equation (\ref{d})
for the parameterized particle and, in view of the relation
$\partial _tT = N$, the Schr\"odinger equation, as written
in a generic time variable reads

\begin{equation}
i\hbar \partial _t\psi = N{\cal H}\psi
\, .
\label{gn}
\end{equation}

This result is completely equivalent to apply, for our
system, the quantum dynamics
proposed in \cite{M02};
Though the considerations developed up to the end of this
sections holds in a generic time variable,
now we come back to a synchronous frame. 

The choice of the normal ordering
$p_R^2/R \rightarrow -\hbar ^2\partial _R(1/R)\partial _R$
is obliged by the requirement to turn the space of the
solutions of equation (\ref{p}) into an Hilbert one.
In fact, when taken
in this form, the Hamiltonian ${\cal H}$ results to be
Hermitian and therefore, by adopting the usual Dirac
\emph{bra-ket} notation, we get

\begin{equation}
\partial _T \left( \langle \psi _1\mid \psi _2 \rangle \right) 
= \frac{i}{\hbar }\left( \langle
\psi _1 {\cal H}\mid \psi _2 \rangle
- \frac{i}{\hbar }\langle
\psi _1 \mid {\cal H}\psi _2 \rangle \right) = 0
\, , 
\label{q}
\end{equation}

Being $\psi _1$ and $\psi _2$ two generic state functions.\\
Indeed we can introduce a probability density,
associated to the state function,
defined as $\rho \equiv \psi ^*\psi$; it is easy to recognize 
that it satisfies a continuity equation of the form
$\partial _T\rho + \partial _aJ^a = 0 ,\; a = R, \pm , \phi$,
being, for instance 

\begin{equation}
J^R = 
\sqrt{\frac{3\pi }{2}}\frac{l_{Pl}^2}{2i} \left( 
\frac{\psi }{R} \partial _R\psi ^* -
\frac{\psi ^*}{R} \partial _R\psi \right) 
\,  
\label{qaa}
\end{equation}

and analogous expressions for
$J^{\pm }$ and $J^{\phi }$.\\
Such a continuity equation,
in view of  the Gauss theorem, as
extended to the configuration space, implies that

\begin{equation}
\partial _T\int_{-\infty }^{\infty }
d^{\pm }\beta d\phi  \int_0^{\infty }dR 
\rho (R, \; \beta _{\pm }\, \phi ) = 0
\, \quad \int_{-\infty }^{\infty }
d^{\pm }\beta d\phi  \int_0^{\infty }dR 
\rho = 1
\, . 
\label{r}
\end{equation}

An important step consists of observing that, 
if we expand the wave function as follows

\begin{equation}
\psi (T, \; R, \; \beta _{\pm }, \phi ) =
\int _{-\infty }^{\infty }d\varepsilon 
C(\varepsilon )\chi (\varepsilon , \; R, \; \beta _{\pm }, \; \phi ) 
exp\{ -\frac{i}{\hbar }\varepsilon (T - T_0)\} 
\, , 
\label{s}
\end{equation}

then we get, from the Schr\'odinger equation (\ref{p}),
the eigenvalues problem

\begin{equation}
\hat{{\cal H}} \chi =
\sqrt{\frac{3\pi }{2}}\left\{ \frac{l_{Pl}^2\hbar }{2} \left[
\partial _R\left( \frac{1}{R} \right) \partial _R - 
\frac{1}{R^3}\left( \partial _+^2 + \partial _-^2 \right)
\right] - 
\frac{3c^2}{8\pi } \partial _{\phi }^2 +
U(R,\; \beta _{\pm },\; \phi ) \right\} \chi = \varepsilon \chi 
\, . 
\label{t}
\end{equation}

Here the function $C(\varepsilon )$ is determined by
assigning the initial wave function
at the instant $T = T_0$, i.e.

\begin{equation}
\psi (T = T_0, \; R, \; \beta _{\pm }, \phi ) \equiv
\psi _0(R \; \beta _{\pm }, \phi ) = 
\int _{-\infty }^{\infty }d\varepsilon 
C(\varepsilon )\chi (\varepsilon , \; R, \; \beta _{\pm }, \; \phi ) 
\label{u}
\end{equation}

where we have taken into account the
orthonormality of the eigenfunctions $\chi$'s.\\
Thus (\ref{t}) shows the following important issue:
\emph{to fix the Gaussian reference
leads to the appearance of a non-zero
total quantum energy of the gravity-``matter'' system}.

\section{Classical limit of the theory}

Now, in order to understand the implications of
this new quantum dynamics on the actual Universe, 
let us take the semiclassical expansion
for the wave function, i.e.

\begin{equation}
\psi = exp\{ \frac{i}{\hbar }\sigma (R, \; \beta _{\pm }, \phi )\}
\, \quad
\sigma = \sigma _0 + \frac{\hbar }{i}\sigma _1 +
\left( \frac{\hbar }{i}\right) ^2\sigma _2 + ... 
\label{v}
\end{equation}

and cutting off it, up to the zero-order of
approximation, then the Schr\"odinger
eigenvalues equation (\ref{t})
rewrites as

\begin{equation}
\frac{l_{Pl}^2}{2\hbar } \left[
-\frac{1}{R} \left( \partial _R \sigma _0\right) ^2 + 
\frac{1}{R^3}\left( \partial _+\sigma _0\right) ^2 +
\left( \partial _-\sigma _0\right) ^2
\right] +
\frac{3c^2}{8\pi } \left( \partial _{\phi }\sigma _0\right) ^2 +
U(R,\; \phi ) - \sqrt{\frac{2}{3\pi }}  \varepsilon = 0
\, . 
\label{w}
\end{equation}

Thus we get an Hamilton-Jacobi (H-J) equation in which
appears a new term,
coming from the, no longer zero, eigenvalue
of the Hamiltonian operator;
indeed, such a term is equivalent to the contribution
given by a non-relativistic dust fluid and it can be    
interpreted as the energy density of the reference frame
it-self (in this sense the non-relativistic nature 
comes from its ``comoving state'').
It is worth noting how the interpretation
of $-\varepsilon$ like density of energy
must overcome the fact that it is not positive definite; 
as a solution to this
problem, we propose the idea that the Universe
spontaneously decays into the state of minimal energy,
which, in general, corresponds to negative values of
$\varepsilon$.
This fact is a consequence of being the
super-Hamiltonian non-positive definite
and it implies that the new term
results into a positive definite energy density
of the reference fluid.

Let us now take the following (general) expansion
of the H-J function 

\begin{equation}
\sigma _0 = \rho _0(R) + P_+\beta _+ + P_-\beta _- +
P _{\phi }\phi 
\label{x}
\end{equation}

being the P's generic constants. 
If we assume that,
near enough to the singularity ($R\rightarrow 0$), 
it is possible to neglect the
potential term $U(R, \phi )$, then expression (\ref{x}) 
reduces the above H-J equation (\ref{w}) to the simple
form

\begin{equation}
\frac{l_{Pl}^2}{2\hbar } \left[
-\frac{1}{R} \left( \frac{d\rho _0}{dR}\right) ^2 +
\frac{P^2}{R^3} \right]
- \sqrt{\frac{2}{3\pi }}  \varepsilon = 0
\, \quad 
P^2 \equiv P_+^2 + P_-^2 + \frac{3c^2P_{\phi }^2}{8\pi } 
\, . 
\label{y}
\end{equation}

Hence, we easily get:

\begin{equation}
\sigma _0 = \int dR\left( \frac{1}{R}\sqrt{P^2 - 
eR^3} \right)
+ P_+\beta _+ + P_-\beta _- +
P _{\phi }\phi 
\, ,
\label{xcx}
\end{equation}

being $e\equiv \sqrt{\frac{2}{3\pi }}
\frac{2\hbar }{l_{Pl}^2}\varepsilon $. Now,
in agreement with the H-J method, 
we differentiate with respect tp the constants $P_i$'s
($i = \pm, \phi $) and
(equating the results to certain constants $C_i$) 
arrive to the following expressions:

\begin{equation}
\beta _{\pm } = \int dR\left(
\frac{P_{\pm }}{R\sqrt{P^2 - eR^3}} \right)
+ C_{\pm } \, \quad 
\phi = \int dR\left( \frac{3c^2P_{\phi }}{8\pi R\sqrt{P^2 - eR^3}} \right)
+ C_{\phi } 
\, ,
\label{ycy}
\end{equation}

Now we explicit these solutions,
respectively in the two asymptotic limits
$R\rightarrow 0$ and $R\rightarrow \infty $, i.e.:

\begin{equation}
R\rightarrow 0 \, \Rightarrow \, 
\beta _{\pm } \sim \frac{P_{\pm }}{P}\ln R + C_{\pm } \, \quad
\, ,
\phi \sim \frac{3c^2P_{\phi }}{8\pi P}\ln R + C_{\phi }
\label{z}
\end{equation}

\begin{equation}
R\rightarrow \infty \, \Rightarrow \, 
\beta _{\pm } \sim -\frac{2P_{\pm }}{3\mid q\mid }
\frac{1}{R^{3/2}} + C_{\pm } \, \quad
\, ,
\phi \sim -\frac{2c^2P_{\phi }}{8\pi \mid q\mid }
\frac{1}{R^{3/2}} + C_{\phi } \, \quad
\, , 
\label{z1a}
\end{equation}

where,
in agreement with the above statement,
we required that $e$ be a negative quantity, i.e.
$-e=q^2$.
On the classical limit, this assumption

Equation (\ref{z}) shows that, near the
singularity in $R=0$, the solution takes, as expected,
a Kasner-like form; in particular,
by setting $\pi _i \equiv P_i/P$ ($i=\pm ,\phi $),
then we have $\sum _i \pi _i^2 = 1$.\\
On the other hand,
equation (\ref{z1a}) implies that, far enough
from the singularity, the anisotropy and
scalar field degrees of freedom are frozen out
of the dynamics and the Universe approaches
an isotropic
(the relic anisotropy, being no longer dynamical,
can be ruled out by redefining the 1-forms
$\sigma ^i$'s)
and scalar-field-free expansion. 

Now, the validity of our interpolating solution
requires the possibility to really neglect
the whole potential term $U(R, \phi )$
near enough to the cosmological singularity.
The presence of the scalar field potential
term is surely crucial to generate
the inflationary scenario,
but, sufficiently close to
the initial ``Big-Bang'', its dynamical
role is expected to be
very limited; in fact,
if we neglect the potential term, 
then, remembering that for early time
$R\sim \sqrt{T} \; \Rightarrow \;
H \equiv (1/R)dR/dT \sim 1/2T$,
we get the classical free field solution
$\phi \propto \ln T$ \cite{BK73}. Hence the kinetic 
term of the field reads 
of order $\mathcal{O}(1/T^2)$; 
therefore, having in the limit toward the
``Big-Bang'' ($T\rightarrow 0$) that 
$T^2V_{{\cal T}(T)}(\phi (T))\; \rightarrow \; 0$,
we can conclude that the Coleman-Weinberg
potential is asymptotically negligible
(this behavior remains valid for almost all
inflationary potentials).\\
Taking into account such classical analysis,
we assume that, during the Planck epoch,
when the Universe
performed its
quantum evolution, the potential
of the scalar field plies
no significant role.

Instead, in order to neglect the
ultrarelativistic energy density and the
Bianchi IX potential, with respect to $\varepsilon$, we
have to require that, respectively, the following two
conditions hold:

\begin{equation}
R\gg \frac{\mu ^2}{\mid \varepsilon \mid } 
\, ,
\label{z2a}
\end{equation}

\begin{equation}
R(V(\beta _{\pm }(R))\ll \frac{l_{Pl}^2\mid \varepsilon \mid }
{\hbar c}
\, ,
\label{z3a}
\end{equation}

Since, as shown in \cite{B00}, for $R\rightarrow 0$ the
term $R\mid V - 1\mid $ approaches zero, because of the
scalar field presence, then the above inequality
(\ref{z3a}) reduces to the simpler one $R\ll R^*$
(being $R^* = R^*(\pi _{\pm}, \varepsilon )$).
 Thus, by
(\ref{z2a}), the validity of our approach is
ensured by the inequality

\begin{equation}
\mu \ll \sqrt{R^*\mid \varepsilon \mid} 
\, .
\label{z4a}
\end{equation}

\section{The generic cosmological solution}

In this section we show how the analysis above
developed can be extended locally to a generic 
inhomogeneous cosmological model
\cite{BKL82} (see also \cite{K93,M95}).
The leading idea in such an upgrading of our
homogeneous picture consists of observing that, near the
singularity, the generic cosmological
solution can be approached as a long-wavelength one.
From a quantum point of view, a long-wavelength evolution
corresponds to neglect the spatial gradients of the dynamical
variables, so reducing the Wheeler superspace to the
direct product of $\infty ^3$ minisuperspaces.\\ 
However, as we will see, from the potential structure
it comes out that the terms containing the spatial
gradients are of higher order near enough to the ``Big-Bang''.
Therefore below we will adopt the long-wavelength
approximation because   
the quantum behavior of the real Universe had to be confined
to the Planck era.\\
Indeed to neglect the potential as a whole corresponds
to take the quantum information as independent
over causally disconnected regions. 

We start by observing how, in the ADM formalism,
the line element of a generic inhomogeneous cosmological
model reads as 

\begin{equation}  
ds^2 = -N^2dt^2 + \gamma _{\alpha \beta }
(dx^{\alpha } + N^{\alpha } dt)(dx^{\beta } +
N^{\beta }dt)
\, , 
\label{xtx1}
\end{equation} 

where the 3-metric tensor $\gamma _{\alpha \beta }$
is provided by

\begin{equation}
\gamma _{\alpha \beta }dx^{\alpha }dx^{\beta } =
R^2(t, x^{\alpha })\left(
e^{2(\beta (t, x^{\alpha })}\right) _{ij} \sigma ^i(
x^{\alpha })
\sigma ^j(x^{\alpha })
\, , 
\label{xtx2}
\end{equation}

Above by $\alpha , \beta = 1,2,3$ we denote the spatial
indices and the 1-forms are now constructed
as $\sigma ^i = l^i_{\alpha }(x^{\alpha })dx^{\alpha }$;
the components of the vectors $l^i_{\alpha }$ correspond to
arbitrary functions of the spatial coordinates and,
therefore no special symmetry is assumed. The generality
of this model implies also
that the shift vector $N^{\alpha }$
can be taken no longer zero. 

Then, by adopting the same parameterization 
(\ref{f}), the gravity-``matter'' action
resembles, in a Gaussian reference ($N=1$ and
$N^{\alpha }= 0$), the form

\begin{equation}
S_{Inh} = \int \left\{ p_R\frac{\partial R}{\partial T}
+ \sum _{r}\left( p_{r}\frac{\partial \beta _{r}}{\partial T}
\right) - H(x^{\alpha })\right\}
d^{3}xdT 
\, .
\label{xtx3}
\end{equation}

Here $r = \pm , \phi $ 
and, by $H(x^i)$, we denote the following point dependent
Hamiltonian term 

\begin{equation}
H(x^{\alpha }) = \frac{4\pi }{3J}\left\{ -\frac{p_R^2}{R} 
+ \frac{1}{R^3} \left( p_{+}^{2} + p_-^{2} + \frac{3}{8\pi }
p_{\phi }^{2} \right) +
{\cal U} \right\} 
\, , 
\label{xtx4}
\end{equation}

where $J\equiv {\bf l}^1\cdot {\bf l}^2\wedge
{\bf l}^3$
(the scalar and vector products are taken by
treating the spatial coordinates as Euclidean ones) 
and the potential term ${\cal U}$ is defined by

\begin{eqnarray}
\label{xtx5}
&{\cal U}& = \frac{3R}{128\pi ^2}
\left\{ a ^2_1(x^{\alpha })e^{-8\beta _+} +
a ^2_2(x^{\alpha })e^{4(\beta _+ + \sqrt{3}\beta _-)} +
a ^2_3(x^{\alpha }) e^{4(\beta - \sqrt{3}\beta _-)} +
W(x^{\alpha }, \; R,\; \beta _{\pm },\;
\partial _{\alpha }R,\; \partial _{\alpha }
\beta _{\pm })
\right\} + \nonumber \\
&\frac{3R^3}{4\pi }&V_{{\cal T}}(\phi )
+ \frac{\mu ^2(x^{\alpha })}{R}
\, . 
\end{eqnarray}

Above, by $a_i$ ($i=1,2,3$), we refer to the space quantities 

\begin{equation}
a_i(x^{\alpha }) \equiv {\bf l}^i\cdot rot {\bf l}^i
\label{xtx6}
\, ,
\end{equation}

where we regard again the operations $\wedge $ and $rot $ in
Euclidean sense. 

To outline the relative behavior in the potential terms, as
the singularity is approached for
$R \rightarrow 0$, let us introduce the new
variables

\begin{eqnarray}
\label{xtx7}
D\equiv R^3 \nonumber \\
H_1\equiv \frac{1}{3} + \frac{\beta _
+ + \sqrt{3}\beta _-}{3\ln R} \nonumber \\ 
H_2\equiv \frac{1}{3}
+ \frac{\beta _+ - \sqrt{3}\beta _-}{3\ln  R} \nonumber \\
H_3\equiv \frac{1}{3}
- \frac{2\beta _+}{3\ln R} \nonumber \\
\sum _i H_i = 1 
\, . 
\end{eqnarray}

Taking into account these definitions, the potential
${\cal U}$ rewrites as follows

\begin{eqnarray}
\label{xtx8}
{\cal U} = \sum _i\left( a_i^2D^{4H_i}\right) + W \\
W\sim \sum _{j\neq k}\mathcal{O}
\left( D^{2(H_j + H_k)} \right) 
\, ;
\end{eqnarray}

Now it is easy to realize that, near the cosmological
singularity
($D\rightarrow 0$), 
the term $W$ becomes negligible.
Indeed this conclusion is supported by the classical behavior
of the spatial gradients, which does not destroy
the feature above outlined (see below for the classical solution
of the model).

Summarizing, asymptotically, the system
is described, in a Gaussian reference,  by the action
(\ref{xtx3}), of which super-Hamiltonian reads as follows:

\begin{eqnarray} 
\label{xtx9}
H(x^{\alpha }) = - \frac{p_R^2}{R}
+ \frac{1}{R^3} \left( p_{+}^{2} + p_-^{2} + \frac{3}{8\pi }
p_{\phi }^{2} \right) + \nonumber \\ 
\frac{3R}{128\pi ^2}
\left\{ a ^2_1(x^{\alpha })e^{-8\beta _+} +
a ^2_2(x^{\alpha })e^{4(\beta _+ + \sqrt{3}\beta _-)} +
a ^2_3(x^{\alpha} ) e^{4(\beta - \sqrt{3}\beta _-)}  
\right\} + \nonumber \\
\frac{3R^3}{4\pi }V(\phi ) + \frac{\mu ^2}{R}
\, . 
\end{eqnarray}

We recognize that, in this asymptotic form,
the dynamics of the generic cosmological solution corresponds
to extend, in each space point, the same evolution discussed
for the Bianchi IX model. 

The absence of spatial gradients of the dynamical variables
(the function $a_i(x^{\alpha })$ specify the considered
inhomogeneous model) implies that, if we parameterize the
Gaussian time as $T=T(t,\; x^{\alpha })$, then the new action
takes the form 

\begin{equation}
S_{Inh} = \int \left\{ p_R\frac{\partial R}{\partial t}
+ \sum _{r}\left( p_{r}\frac{\partial \beta _{r}}{\partial t}
\right) + p_T\frac{\partial T}{\partial t} - 
\Lambda (t, x^{\alpha })\left(p_T 
+ H(x^{\alpha })\right) \right\}
d^{3}xdt 
\, . 
\label{xtx10}
\end{equation}

Above we adopted the same procedure developed for the
homogeneous case in section $3$; here $p_T$ plays the role
of conjugate momentum to the Gaussian time, while $\Lambda $
denotes a Lagrangian multiplier for which takes place the
relation

\begin{equation}
\Lambda (t, x^{\alpha }) =
\frac{\partial T}{\partial t} = N(t, x^{\alpha })
\, .
\label{xtx11}
\end{equation}

The canonical quantization of this dynamical system
is achieved by implementing on operator level the
super-Hamiltonian constraint, i.e. we have to require that it
annihilates the state functional
$\Psi (T(x^{\alpha }),\; R(x^{\alpha }),\; 
\beta _{\pm }(x^{\alpha }), \; \phi (x^{\alpha }))$;
therefore, the quantum dynamics is described by the
following  $\infty ^3$ Schr\"odinger equations

\begin{equation}
i\hbar \frac{\delta \Psi }{\delta T(x^{\alpha })} = 
\hat{H}(x^{\alpha })\Psi
\, .
\label{xtx12}
\end{equation} 

where now the momentum operators are expressed in terms of
functional derivatives (instead of ordinary ones like in the
homogeneous case). 

If we introduce the definition
$\partial _t(\; ) \equiv
\int_{\Sigma^3_t}(\delta \Psi /\delta T)\partial _tTd^3x$
(being $\Sigma ^3_t$ the one parameter family of spatial
hypersurfaces filling the space-time), then,
taking into account the relation (\ref{xtx11}),
the above system of $\infty ^3$ (independent) equations
(\ref{xtx12}) can be smeared as follows 

\begin{equation}
i\hbar \partial_t \Psi = \hat{{\cal H}}\Psi \equiv 
\int _{\Sigma ^3_t}d^3x \left(
N\hat{H}(x^{\alpha })\right) \Psi 
\, .
\label{xtx13}
\end{equation}

Since here the lapse function must be regarded as assigned, then
$T$ is known by (\ref{xtx11}) and the wave functional depends
directly on the label time $t$, i.e.
$\Psi = \Psi (t, \; R, \; \beta _{\pm }, \; \phi )$. 

It remains to require that the wave functional $\Psi$
be invariant under space diffeomorphisms,
i. e. transformations of the form
$x^{\alpha }\rightarrow x^{\alpha } + \xi ^{\alpha }$, where
$\xi ^{\alpha }$ denotes a generic infinitesimal displacement;
as effect of this transformation $\Psi $ changes by the
amount:

\begin{equation}
\delta \Psi = \int _{\Sigma ^3_t}d^3x
\left\{ \left[ 
\frac{\delta \Psi }{\delta R}\partial _{\alpha }R +
\frac{\delta \Psi }{\delta \beta _+}\partial _{\alpha }\beta _+
+ \frac{\delta \Psi }{\delta \beta _-}\partial _{\alpha }
\beta _- +
\frac{\delta \Psi }{\delta \phi }\partial _{\alpha }\phi
\right] \xi ^{\alpha }\right\} 
\, . 
\label{xtx14}
\end{equation}

Since $\xi ^{\alpha }$ is a generic 3-vector, then
$\delta \Psi$
vanishes only if the following equation holds

\begin{equation}
\frac{\delta \Psi }{\delta R}\partial _{\alpha }R +
\frac{\delta \Psi }{\delta \beta _+}\partial _{\alpha }\beta _+
+ \frac{\delta \Psi }{\delta \beta _-}\partial _{\alpha }
\beta _- +
\frac{\delta \Psi }{\delta \phi }\partial _{\alpha }\phi = 0 
\, . 
\label{xtx15}
\end{equation}

This equation corresponds to the super-momentum constraint
which has to appear because of the transformation
$T = T(t, x^{\alpha })$. 

It is worth noting how equations (\ref{xtx13})
and {\ref{xtx15}) coincides with the implementation to the
present case of the analysis developed in \cite{M02}.
The general nature of our cosmological model makes 
such a coincidence of physical interest; in fact,
either the present analysis, as well as that one outlined in
\cite{M02}, lead to the issue that
\emph{to fix a reference frame before quantizing the metric
field implies that a time evolution is restored in the
dynamics}. 

Equation (\ref{xtx13}) is reduced to an eigenvalues problem,
as soon as we expand the wave functional in the form

\begin{equation}
\Psi = \int _{{\cal F}} D\varrho
\chi (\varrho ,\: R,\; \beta _{\pm }\; ,\phi )
exp \left\{ - \frac{i}{\hbar } \int _{t_0}^tdt 
\int _{\Sigma ^3_t}d^3x
N(t, x^{\alpha })\varrho (x^{\alpha })\right\} 
\, , 
\label{xtx16}
\end{equation}

where $D\varrho $ denotes the Lebesgue measure on the
functional space ${\cal F}$; in fact, substituting (\ref{xtx16})
into (\ref{xtx13}), we get

\begin{equation}
\hat{H}(x^{\alpha })\chi = \varrho (x^{\alpha }) \chi
\, . 
\label{xtx18}
\end{equation}

For the generic cosmological solution,
this equation explicitly reads

\begin{equation}
\left[ \hbar ^2 \frac{\delta }{\delta R}
\frac{1}{R}\frac{\delta }{\delta  R} 
- \frac{\hbar ^2}{R^3} \left(
\frac{\delta ^2}{\delta \beta _+^2} +
\frac{\delta ^2}{\delta \beta _-^2} +
+ \frac{3\hbar ^2}{8\pi } 
\frac{\delta ^2}{\delta \phi ^2} 
\right) + {\cal U} \right] \Psi = \varrho \Psi
\, .
\label{xtx19}
\end{equation}

Point by point in space, the above equation resembles the
corresponding one for the Bianchi IX model, i.e. (\ref{t});
therefore all the results following in Section $3$ and $4$
hold as extended to the inhomogeneous case. 
In particular, it is worth noting that, even for the generic
case, to quantize the system in a Gaussian frame,
endow the quantum dynamics with a time evolution and
allow to define an Hilbert space for the states of the theory.
Furthermore, it can be shown, along the same lines of
section $4$,  that in the semiclassical limit
$\varrho (x^{\alpha })$ induces the Universe isotropization.
In fact, the ground state of the Universe has to correspond 
everywhere to a negative energy eigenvalue
(i.e. $\forall x^{\alpha }\, :\, \varrho < 0$), because of the
super-Hamiltonian is not positive definite;
such a ground state, if it is stable, 
(i.e. the negative spectrum of energies is bounded by below),
corresponds, in the classical limit, to the phenomenology of
a dust fluid filling the Universe and having energy density
$-\rho /R^3$.

We conclude this section by showing how the model here
considered corresponds, in the classical limit, to a generic
inhomogeneous model. We recall that the pure
gravitational field, in the general picture, 
requires to be described four physically arbitrary functions
of the spatial coordinates. 

Once taken the following expansion for the
wave functional

\begin{equation}
\Psi = exp\{ \frac{i}{\hbar }\Sigma (R, \; \beta _{\pm }, \phi )\}
\, \quad
\Sigma = \Sigma _0 + \frac{\hbar }{i}\Sigma _1 +
\left( \frac{\hbar }{i}\right) ^2\Sigma _2 + ... 
\, , 
\label{vxt}
\end{equation}

in the limit $\hbar \rightarrow 0$, the zero order of
approximation, to the eigenvalue problem (\ref{xtx19})
and to the super-momentum constraint (\ref{xtx15}), yields the
following Hamilton-Jacobi equations 

\begin{equation}
-\frac{1}{R}\left( \frac{\delta \Sigma _0}{\delta R}\right) ^2
+ \frac{1}{R^3} \left[
\left( \frac{\delta \Sigma _0}{\delta \beta _+}\right) ^2 +
\left( \frac{\delta \Sigma _0}{\delta \beta _-}\right) ^2
\right] + \frac{3}{8\pi }
\left( \frac{\delta \Sigma _0}{\delta \phi }\right)^2 
+ {\cal U} - \varrho = 0 
\, . 
\label{xtx20}
\end{equation}

\begin{equation}
\frac{\delta \Sigma _0}{\delta R}\partial _{\alpha }R +
\frac{\delta \Sigma _0}{\delta \beta _+}\partial _{\alpha }\beta _+
+ \frac{\delta \Sigma _0 }{\delta \beta _-}\partial _{\alpha }
\beta _- +
\frac{\delta \Sigma _0}{\delta \phi }\partial _{\alpha }\phi
= 0
\, . 
\label{xtx21}
\end{equation}

In the asymptotic limit toward the cosmological singularity
($R\rightarrow 0$), the solution to these equations
reads, for an expanding Universe, as

\begin{equation}
\Sigma _0 = \int _{\Sigma ^3_t}\left( 
-K(x^{\alpha })\ln R +
K_+(x^{\alpha })\beta _+ +
K_-(x^{\alpha })\beta _- +
K_{\phi }(x^{\alpha })\phi \right) d^3x
\, ,
\label{xtx21a}
\end{equation}

with

\begin{equation}
K^2\equiv 
K_+^2 + K_-^2 + \frac{3}{8\pi }K_{\phi }^2 
\, . 
\label{xtx22}
\end{equation}

Indeed if we take the functional derivatives of $\Sigma _0$
with respect to the $K$'s and equate them to space functions,
then the anisotropies and the scalar field acquire the
following dependence on $R$:

\begin{equation}
\beta _{\pm }(R) = -\frac{K_{\pm }}{K}\ln R +
\beta ^*_{\pm }(x^{\alpha })
\, \quad
\phi = -\frac{K_{\phi }}{K}\ln R +
\phi  ^*(x^{\alpha }) 
\, .
\label{xtx23}
\end{equation}

or, by an obvious position, the above expressions rewrite as

\begin{equation}
\beta _{\pm }(R) = -\Pi _{\pm }\ln R +
\beta ^*_{\pm }(x^{\alpha })
\, \quad
\phi = -\Pi _{\phi }\ln R +
\phi  ^*(x^{\alpha }) 
\, .
\label{xtx24}
\end{equation}

By (\ref{xtx23}),
once taken into account the additional constraint

\begin{equation} 
\Pi _+^2(x^{\alpha }) + \Pi _-^2(x^{\alpha }) +
\Pi _{\phi }^2(x^{\alpha }) = 1
\, ,
\label{pi}
\end{equation}

then 
equation (\ref{xtx20}) is
automatically satisfied by (\ref{xtx24}), while the validity
of equation (\ref{xtx21}) implies that 

\begin{equation}
\Pi _+ \partial _{\alpha }\beta ^*_+ +
\Pi _- \partial _{\alpha }\beta ^*_- + 
\Pi _{\phi } \partial _{\alpha }\phi ^* = 0 
\, .
\label{xtx25}
\end{equation}

In the considered approximation and in view of
solution (\ref{xtx24}), 
the potential ${\cal U}$ is of higher order with respect to
the retained terms which behave like
$\mathcal{O}(1/R^3)$.
The functions $\beta ^*_{\pm }$ do not
correspond to real new degrees of freedom because, by the
metric tensor (\ref{xtx2}), we see that they can be
included in the definition of the vectors $l^i_{\alpha }$;
in this respect, we can think of these functions
like two of the nine components
of the 1-forms vectors.\\
Since the six functions $\Pi _{\pm }\; ,
\Pi _{\phi }\; , \beta ^*_{\pm }\; ,
\phi ^*$ have to satisfy the four equations (\ref{pi}) and
(\ref{xtx25}), only two of them are really available
for the Cauchy problem; adding these two free functions
to the remaining seven components of the vectors
$l^i_{\alpha }$ (two components were identified with
$\beta ^*_{\pm }$), we arrive to nine independent functions. 
However, taking into account the space
diffeomorphisms
$x^{\alpha ^{\prime }} =
x^{\alpha ^{\prime }} (x^{\alpha })$ to kill other three
degrees of freedom, we see that our solution contains the right
number of physically arbitrary functions of spatial coordinates
associated to the generality. In fact four functions
correspond to the generic gravitational field and the
remaining two are available for a generic scalar field;
finally we observe that the functions $\mu $ and $\varrho $
are not affected by any restriction.
These considerations show how our model describes,
on a classical level, a generic inhomogeneous cosmology and
therefore, some of the features above outlined by
its quantum dynamics have general validity.\\ 
Indeed 
the reduction of the
superspace into the direct product of
$\infty ^3$ minisuperspaces,
each for each space point, is based on the long-wavelength
approximation and therefore it is due to the structure
of the Einstein theory near the cosmological singularity;
but the appearance of a non-zero eigenvalue of the
super-Hamiltonian as a consequence of the frame fixing
constitutes a general picture.

\section{Concluding remarks}

In order to summarize the outcomes of this work, we can
arrange them into the following main points.

i) By extending to the minisuperspace the same method
for quantizing, via an Hamiltonian constraint,
the non-relativistic particle, we could perform the
canonical quantization of a Bianchi type IX model
(in presence of ultrarelativistic
matter and a real self-interacting scalar field),
as viewed in a Gaussian reference frame.
The issue of such a procedure
consists of a quantum dynamics having
a Schr\"odinger
morphology, an appropriate Hilbert space
of its states and
a corresponding notion of probability
density for the system configuration.\\
The main difference between such a canonical quantization
in a fixed frame and the Wheeler-DeWitt one,
consists of the super-Hamiltonian spectrum;
in fact the ``frame fixing'' removes the
time displacements invariance of the theory and 
restores non-zero super-Hamiltonian eigenvalues.

ii) When taking
the semiclassical limit of this quantum theory, 
via a standard WKB approach, we get the H-J
equation corresponding to the considered model, but with
the appearance of a new energy contribution, which
reflects the classical outcoming of the no longer zero
eigenvalue of the super-Hamiltonian. If we argue that the
Universe is forced to approach the (quantum) state
of minimal energy and observe that the
super-Hamiltonian is non-positive definite, then
it is natural that the system settle down, in 
the classical limit, into a state of negative value of $\varepsilon$.

iii) The negative nature of $\varepsilon$ implies
that the new term appearing in the dynamics resembles
a non-relativistic matter component 
and therefore contributes to
the total critical parameter of the Universe 
as ``dark matter'' component
does.\\
Furthermore we have shown how 
this non-relativistic matter,
under rather general conditions,
becomes dominant and drives    
a process of isotropization which brings the Universe
from a Kasner-like regime
to a stage of frozen anisotropies and scalar field. 

For a detailed discussion on the cosmological
implications of such additional dark matter component,
in the case of a closed Robertson-Walker model
(i. e. the actual one with $\beta _+ = \beta _- = 0$),
see \cite{M03}.

iv) Near the cosmological singularity, the minisuperspace
picture above outlined can be extended, point by point
in space,
to the generic cosmological solution; From a physical point of
view this result is equivalent to claim that, within each
causal region, the quantum behavior is completely equivalent
to that one proper of a Bianchi IX model; therefore
all the results, as described by the above three points,
hold locally for a generic inhomogeneous cosmology.\\
In the quantum regime, the reduction of the superspace
to the direct
product of $\infty ^3$ minisuperspaces, corresponds
to adopting the long-wavelength approximation, i.e.
to neglect the spatial gradients of the dynamical variables.
We stress, in this respect, that the 
long-wavelength feature is dynamically induced by the
asymptotic classical evolution and therefore
we argue that it 
survives in the quantum dynamics which       
takes place just during the Planckian era toward
the initial singularity.

\end{document}